\documentclass[aip,jcp,floatfix,amsmath,amssymb,reprint,longbibliography]{revtex4-1}

\usepackage{amsfonts,amstext,amsgen,amsbsy,amsopn,amssymb}
\usepackage{graphicx}
\usepackage{multirow}
\usepackage{siunitx}  

\newcommand{\cm}{\ensuremath{\mathrm{cm}^{-1}}}
\newcommand{\half}{\ensuremath{\frac{1}{2}}}

\begin{document}

\date{\today}

\author{Steven E. Strong}
\thanks{S.E.S and N.J.H contributed equally to this work}
\affiliation{Pritzker School of Molecular Engineering, The University of Chicago, Chicago,
Illinois 60637, USA}

\author{Nicholas J. Hestand}
\thanks{S.E.S and N.J.H contributed equally to this work}
\affiliation{Pritzker School of Molecular Engineering, The University of Chicago, Chicago,
Illinois 60637, USA}
\affiliation{Department of Natural and Applied Sciences, Evangel University, Springfield, Missouri 65802, USA}

\title{Modeling Nonlocal Electron-Phonon Coupling in Organic Crystals Using Interpolative Maps: The Spectroscopy of Crystalline Pentacene and 7,8,15,16-Tetraazaterrylene}

\begin{abstract}
Electron-phonon coupling plays a central role in the transport properties and photophysics of organic crystals.
Successful models describing charge- and energy-transport in these systems routinely include these effects.
Most models for describing photophysics, on the other hand, only incorporate local electron-phonon coupling to intramolecular vibrational modes, while nonlocal electron-phonon coupling is neglected.
One might expect nonlocal coupling to have an important effect on the photophysics of organic crystals, because it gives rise to large fluctuation in the charge-transfer couplings, and charge-transfer couplings play an important role in the spectroscopy of many organic crystals.
Here, we study the effects of nonlocal coupling on the absorption spectrum of crystalline pentacene and 7,8,15,16-tetraazaterrylene.
To this end, we develop a new mixed quantum-classical approach for including nonlocal coupling into spectroscopic and transport models for organic crystals.
Importantly, our approach does not assume that the nonlocal coupling is linear, in contrast to most modern charge-transport models.
We find that the nonlocal coupling broadens the absorption spectrum non-uniformly across the absorption line shape.
In pentacene, for example, our model predicts that the lower Davydov component broadens considerably more than the upper Davydov component, explaining the origin of this experimental observation for the first time.
By studying a simple dimer model, we are able to attribute this selective broadening to correlations between the fluctuations of the charge-transfer couplings.
Overall, our method incorporates nonlocal electron-phonon coupling into spectroscopic and transport models with computational efficiency, generalizability to a wide range of organic crystals, and without any assumption of linearity.
\end{abstract}

\maketitle

\section{Introduction\label{sec:intro}} 
Absorption and photoluminescence spectroscopies are powerful techniques for interrogating the electronic structure of conjugated organic materials.
Spectral shifts, vibrionic peak ratio changes, and line broadening all provide information about the sign and magnitude of the exciton coupling, the curvature and width of the exciton band, the exciton coherence length, and the disorder within the system.\cite{spano_spectral_2010,hestand_expanded_2018} 
Over the past several decades, significant effort has been devoted to develop a comprehensive understanding of these spectroscopic signatures, with a great deal of success.\cite{tretiak_density_2002,spano_spectral_2010,spano_h_2014,hestand_molecular_2017,hestand_expanded_2018,nelson_non_2020,koehler_electronic_2015,bredas_wspc_2016}

One of the main challenges in modeling conjugated organic systems is the importance of electron-phonon coupling,\cite{ostroverkhova_organic_2016,bredas_charge_2004,fratini_transient_2016,bredas_wspc_2016,coropceanu_charge_2007} 
which is commonly divided into two types: local and nonlocal.\cite{coropceanu_charge_2007}
Local electron-phonon coupling is the modulation of the electronic Hamiltonian by predominately intramolecular phonons, while nonlocal~(Peierls) electron-phonon coupling is the modulation of the electronic Hamiltonian by predominately lattice phonons.
Both types are important for accurate descriptions of excited states in organic systems, and several model Hamiltonians have been devised to account for these effects: Holstein models describe local electron-phonon coupling,\cite{holstein_studies_1959,holstein_studies_1959-1} Su-Schrieffer-Heeger models describe nonlocal electron-phonon couplings,\cite{su_solitons_1979} and extended Holstein models describe both simultaneously.\cite{lee_vibronic_2017,fetherolf_unification_2020,duan_ultrafast_2019}
In terms of spectroscopy, local vibronic coupling is responsible for the pronounced $\sim$1400 \cm{} vibronic progression observed in the optical response of many conjugated organic systems and is routinely incorporated in spectroscopic models.
However, far less attention has been paid to the role that nonlocal electron-phonon coupling plays in the optical response.

Several works have shown that nonlocal electron-phonon coupling gives rise to large fluctuations in the charge-transfer~(CT) interactions within organic systems, on the order of their average values.
\cite{troisi_electronic_2005,troisi_dynamics_2006,arago_dynamics_2015,arago_regimes_2016,wang_multiscale_2010} 
This phenomenon arises from the facts that 1) optical lattice phonons have energies well below $k_B T$ at room temperature\cite{dellavalle_intramolecular_2004}~(in contrast with inorganic systems\cite{kulda_inelastic_1994}), and 2) the CT couplings are very sensitive to molecular packing; displacements on the order of the carbon-carbon bond length can dramatically alter the magnitude \textit{and} sign of these quantities.\cite{kazmaier_theoretical_1994,gisslen_crystallochromy_2009} 

The majority of work considering nonlocal electron-phonon coupling has focused on its important role in charge transport, where it results in charge carrier localization and limits carrier mobility.\cite{troisi_charge-transport_2006,troisi_prediction_2007,coropceanu_charge_2007,wang_multiscale_2010,troisi_charge_2011,wang_mixed_2011,ciuchi_transient_2011,fratini_transient_2016,schweicher_chasing_2019} 
The same CT couplings that are perturbed by the nonlocal electron-phonon coupling, however, also play an important role in the spectroscopic response of many organic systems.
In closely packed organic crystals, for example, the optically bright Frenkel excitons couple through a short-ranged, CT mediated ``superexchange'' mechanism.\cite{harcourt_rate_1994,scholes_rate_1995} 
This implies that the spectroscopy of organic crystals should also be sensitive to fluctuations in the CT couplings.\cite{klugkist_scaling_2008,fidder_optical_1991,arago_dynamics_2015,arago_regimes_2016,fornari_exciton_2016}
Surprisingly, however, absorption spectra of many organic crystals can be successfully modeled without including nonlocal electron-phonon coupling~(Ref.~\citenum{hestand_expanded_2018} and citations within).

To resolve this apparent discrepancy, we develop a new method for modeling the spectroscopy of organic crystals that incorporates nonlocal electron-phonon coupling, in addition to the other common ingredients of spectroscopic models: Frenkel excitons, CT excitons, and local electron-phonon coupling.
The method is based on a mixed quantum-classical approach that treats the low-frequency phonon modes classically via molecular dynamics~(MD) simulations while the high-frequency intramolecular vibrations and electronic degrees of freedom are treated quantum-mechanically using a Holstein-style Hamiltonian.
We parameterize the Hamiltonian at each time step according to the MD trajectory, using a mapping approach to make repeated evaluation of the CT couplings computationally tractable.
Both the mixed quantum-classical approach and the map are similar in spirit to approaches used to model the vibrational spectroscopy of condensed phases.\cite{baiz_vibrational_2020}
Importantly, our approach allows for the treatment of nonlocal electron-phonon couplings very generically, and can account for arbitrarily complex dependence of the couplings on the intermolecular structure.
In particular, it is not necessary to make the common assumption that the nonlocal electron-phonon coupling is linear.

\begin{figure*}[t]
\includegraphics{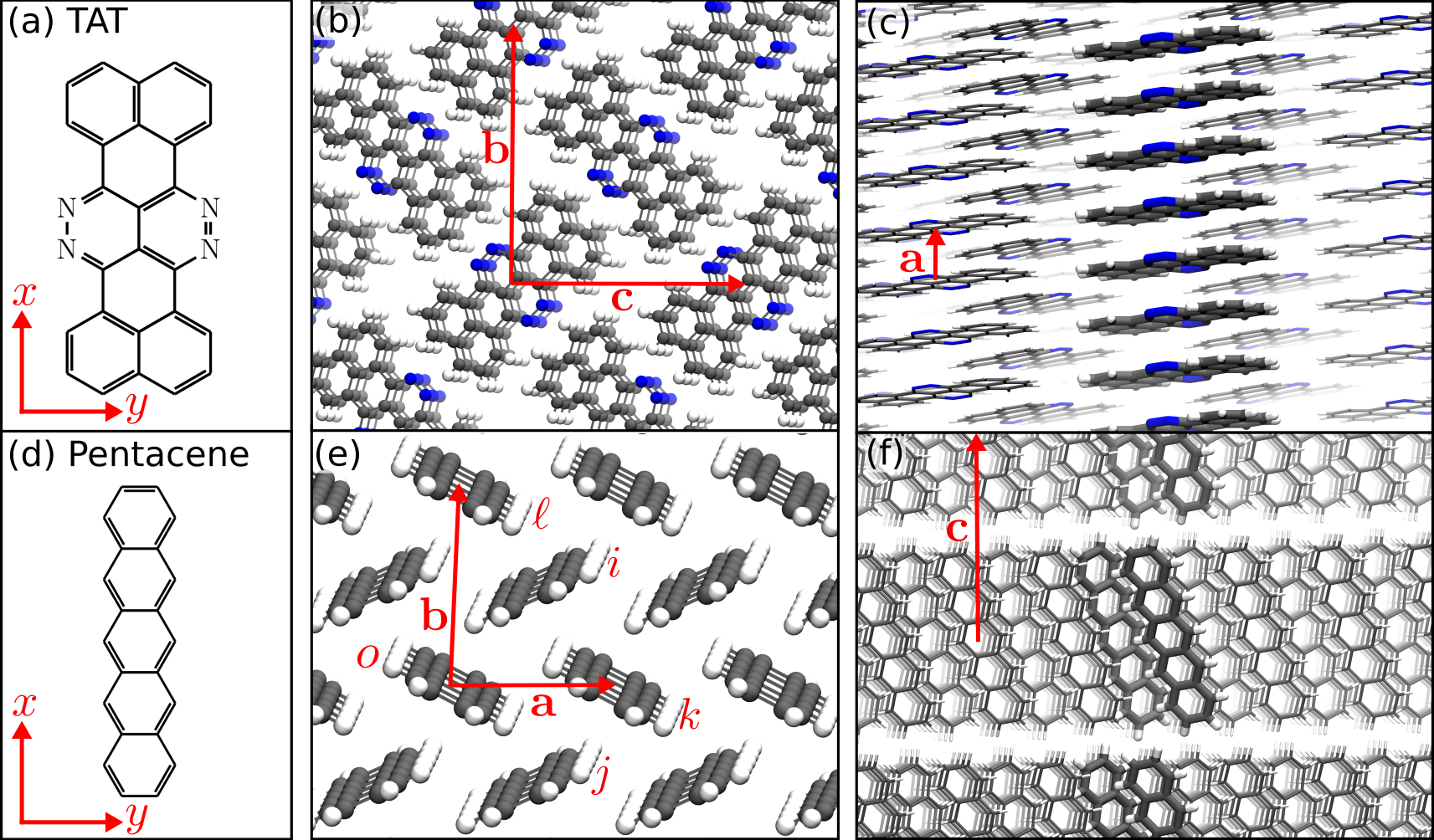}
\caption{TAT~(a) forms pillars in its crystalline state~(b,c).\cite{fan_synthesis_2012}
Panel (b) is a view down these pillars, roughly along the lattice $\mathbf{a}$-direction.
Panel (c) highlights the molecules in one pillar.
Pentacene~(d) forms a crystal with a herringbone structure in the crystallographic $\mathbf{ab}$-plane~(e) and a layered structure in the $\mathbf{c}$-direction~(f).\cite{holmes_nature_1999}
The two molecules in one unit cell and their periodic images along the lattice $\mathbf{c}$-direction are highlighted in panel (f) to guide the eye.
The $x$- and $y$-axes in panels (a) and (d) define the molecular coordinate system used to describe the geometry of a pair of molecules in our map.
The axes in panels (b,c,e,f) are the crystalline axes.
The molecules labelled in panel (e) illustrate the nomenclature we use to classify the four different types of neighbors in pentacene: $\mathbf{r}_{oi}= \pm(\half,\half,0)$, $\mathbf{r}_{oj}= \pm(\half,-\half,0)$, $\mathbf{r}_{ok}= \pm(1,0,0)$, and $\mathbf{r}_{o\ell}= \pm(0,1,0)$.
}
\label{fig:schematic}
\end{figure*}

We apply our method to understand the effects of nonlocal electron-phonon coupling on the absorption spectroscopy of organic crystals.
We focus on two specific systems that exhibit different packing motifs: pentacene, which packs in a herringbone structure,\cite{holmes_nature_1999} and 7-8-15-16-tetraazaterrylene (TAT), which exhibits a slipped $\pi$-stacking structure~(Fig.~\ref{fig:schematic}).\cite{fan_synthesis_2012,wise_spectroscopy_2014}
We chose these systems for four reasons:
First, because CT couplings are known to play an important role in the spectroscopy of both systems, making them susceptible to nonlocal electron-phonon couplings like those we consider here.\cite{yamagata_hj_2014,hestand_polarized_2015}
Second, to evaluate the effects of nonlocal coupling between systems with different packing motifs and therefore different nonlocal coupling forms and strengths.
Third, the absorption spectroscopy of both systems has previously been modeled in the absence of nonlocal coupling, so significant parameterization efforts are not required.\cite{yamagata_hj_2014,hestand_polarized_2015}
And finally, because these and related systems have received considerable attention as promising organic semiconductors.\cite{koehler_electronic_2015,bredas_wspc_2016,ostroverkhova_organic_2016,schweicher_molecular_2020}
We find that both systems exhibit significant fluctuations in the CT and total excitonic coupling due to nonlocal electron-phonon coupling, in agreement with previous results.\cite{troisi_electronic_2005,troisi_dynamics_2006,arago_dynamics_2015,arago_regimes_2016,wang_multiscale_2010} 
These fluctuations broaden the absorption spectrum, but interestingly, the broadening is not uniform.
We find that this nonuniformity is due at least in part to correlations in the electron- and hole-transfer couplings.
This is most obvious for pentacene where the $||\mathbf{a}$-polarized lower Davydov component broadens six times more than the ${\perp}\mathbf{a}$-polarized upper Davydov component.

\section{Methods}
\subsection{Model Hamiltonian\label{sec:model}}
Our model is motivated by a separation in energy scales.
In organic crystals, the intermolecular degrees of freedom oscillate at much lower frequencies than the intramolecular and electronic degrees of freedom due to the weak van der Waals forces that hold the crystals together.
With this in mind, we separate the Hamiltonian into a classical part $H_\mathrm{cls}$ that accounts for low-frequency intermolecular vibrations, and a quantum mechanical part $\hat H_\mathrm{qm}$ that accounts for the high-frequency degrees of freedom.\cite{skinner_vibrational_2009,wang_mixed_2011,shi_modeling_2018,cerezo_adiabatic-molecular_2019}

The quantum-mechanical part of the Hamiltonian is Holstein-like and includes Frenkel excitons, CT excitons, and vibronic coupling
\begin{equation}
    \hat{H}_\mathrm{qm}=\hat{H}_\mathrm{FE}+\hat{H}_\mathrm{CT}+\hat{H}_\mathrm{vib}.
\label{eq:ham}
\end{equation}
Here $\hat{H}_\mathrm{FE}$ represents the Frenkel exciton Hamiltonian
\begin{equation}\label{eq:ham_fe}
    \hat{H}_\mathrm{FE}=\sum_{i}\left(E_{S_1}+\Delta_{0-0}\right)B_{i}^{\dagger}B_{i}
    +\sum_{i\ne j} J\left(\mathbf{q}^{N}_{i},\mathbf{q}^{N}_{j}\right)B_{i}^{\dagger}B_{j}
\end{equation}
where $i$ and $j$ label the molecules in the crystal
and the operator $B_i^{\dagger}$ ($B_{i}$) creates (annihilates) a Frenkel exciton on molecule $i$.
The first sum in $\hat{H}_\mathrm{FE}$ accounts for the energy of a localized Frenkel exciton; $E_{S_{1}}$ is the excitation energy for the $\mathrm{S}_{1}\leftarrow \mathrm{S}_{0}$ transition for a molecule in dilute solution and $\Delta_{0-0}$ is the solution-to-crystal shift that accounts for non-resonant interactions within the crystal.
The second term in $\hat{H}_\mathrm{FE}$ accounts for long-range Coulombic coupling between excitons at sites $i$ and $j$, $J(\mathbf{q}^{N}_i,\mathbf{q}^{N}_j)$, where $\mathbf{q}^{N}_{i}$ is the $3N$ dimensional vector of the atomic coordinates of molecule $i$, and $N$ is the number of atoms in molecule $i$.

The second term in Eq. \ref{eq:ham} is the CT Hamiltonian
\begin{equation}\label{eq:ham_ct}
\begin{split}
    \hat{H}_\mathrm{CT}&=\sum_{i\ne j}E_\mathrm{CT}\left(\mathbf{q}^{N}_{i},\mathbf{q}^{N}_{j}\right)c_i^{\dagger}c_i d_j^{\dagger}d_j\\
    &+\sum_{i\ne j}t_{e}\left(\mathbf{q}^{N}_{i},\mathbf{q}^{N}_{j}\right)c_i^{\dagger}c_j
    +\sum_{i\ne j}t_{h}\left(\mathbf{q}^{N}_{i},\mathbf{q}^{N}_{j}\right)d_i^{\dagger}d_j,
\end{split}
\end{equation}
where the operator $c_{i}^{\dagger}$ ($c_{i}$) creates (annihilates) an electron on molecule $i$ and the operator $d_{i}^{\dagger}$ ($d_{i}$) creates (annihilates) a hole on molecule $i$.
The energy of a CT exciton with the electron localized to molecule $i$ and the hole localized to molecule $j$ is represented by
\begin{equation}\label{eq:ect}
    E_\mathrm{CT}\left(\mathbf{q}^{N}_i,\mathbf{q}^{N}_j\right)=I_{P}-E_{A}+P+V\left(\mathbf{q}^{N}_i,\mathbf{q}^{N}_j\right).
\end{equation}
This energy depends on the atomic positions of the host chromophores $i$ and $j$ through the Coulomb binding energy $V\left(\mathbf{q}^{N}_i,\mathbf{q}^{N}_j\right)$.
The ionization potential $I_P$, electron affinity $E_A$, and polarization energy $P$ also enter into the expression for $E_\mathrm{CT}$ but we approximate these terms to be independent of the coordinates of the host molecules.
The second and third sums in $\hat{H}_\mathrm{CT}$ account for short-range CT coupling with $t_e\left(\mathbf{q}^{N}_i,\mathbf{q}^{N}_j\right)$ and $t_h\left(\mathbf{q}^{N}_i,\mathbf{q}^{N}_j\right)$ representing the electron- and hole-transfer couplings between sites $i$ and $j$.
These terms couple the Frenkel and CT states, and control charge hopping within the CT manifold.

The third term in Eq.~\ref{eq:ham} accounts for the high-frequency intramolecular vibrations responsible for the vibronic progression observed in the absorption spectrum
\begin{equation}\label{eq:ham_vib}
\begin{split}
    \hat{H}_\mathrm{vib}&=\hbar\omega_\mathrm{vib}\sum_{i}b_{i}^{\dagger}b_{i}
    +\hbar\omega_\mathrm{vib}\sum_{i}\left\{\lambda\left(b_{i}^{\dagger}+b_{i}\right)+\lambda^2\right\}B_{i}^{\dagger}B_{i}\\
    &+\hbar\omega_\mathrm{vib}\sum_{i\ne j}\left\{\lambda_{e}\left(b_{i}^{\dagger}+b_{i}\right)+\lambda_{h}\left(b_{j}^{\dagger}+b_{j}\right)+\lambda_{h}^{2}+\lambda_{e}^{2}\right\}c_i^{\dagger}c_i d_j^{\dagger}d_j.
\end{split}
\end{equation}
Here the operator $b_{i}^{\dagger}$ ($b_{i}$) creates (annihilates) a vibrational quantum on chromophore $i$ with energy $\hbar\omega_\mathrm{vib}$.
The first summation therefore describes the vibrational energy of each molecule.
The final three terms account for the local electron-phonon coupling of the exciton, electron, and hole to the intramolecular vibration.
The Huang-Rhys factors $\lambda^2$, $\lambda_e^2$, and $\lambda_h^2$, describe the shift in the nuclear potential relative to the ground state when the molecule hosts a Frenkel exciton, electron, or hole, respectively.
In principle, all intramolecular vibrational modes can be accounted for, but this is prohibitively expensive for all but the simplest molecules.
Instead, we treat the numerous closely spaced modes that contribute to the vibronic progression empirically using a line width that depends on the number of vibrational quanta in the absorbing state, as discussed by Yamagata et al.\cite{yamagata_hj_2014}

Nonlocal electron-phonon coupling enters the Hamiltonian classically through the time-dependent fluctuations in the atomic coordinates.
That is, in Eqs.~\ref{eq:ham}--\ref{eq:ham_vib}, $\mathbf{q}^{N}_{i}\rightarrow\mathbf{q}^{N}_{i}(t)$.
We model these fluctuations using MD simulations of the crystal.
As discussed above, this corresponds to a separation of the Hamiltonian into the quantum mechanical one given in Eq.~\ref{eq:ham} and the classical Hamiltonian
\begin{equation}\label{eq:classicalHam}
    H_\mathrm{cls} = \sum_\alpha \frac{\mathbf{p}_\alpha^2}{2m_\alpha} + U(\{\mathbf{q}\})
\end{equation}
where $\mathbf{p}_\alpha$ is the momentum conjugate to atomic coordinate $\mathbf{q}_\alpha$, $m_\alpha$ is the mass of particle $\alpha$, and $U(\{\mathbf{q}\})$ is the potential used in the MD simulation~(Sec.~\ref{sec:md}), which is, in principle, a function of the set of all the coordinates $\{\mathbf{q}\}$.
Here the sum over $\alpha$ goes over atoms, not molecules.
We then use various ab-initio techniques and the mapping approach developed in Section~\ref{sec:TATmap} to compute the $\mathbf{q}^{N}_{i}$-dependent quantities in the Hamiltonian~(Eq.~\ref{eq:ham}).
If the coordinates $\{\mathbf{q}\}$ are instead static and taken from an experimental crystal structure, the quantum mechanical Hamiltonian reduces to the time-independent one used in previous work to model the optical properties of many different organic crystals.\cite{hestand_expanded_2018}

In reality, all of the quantities appearing in the quantum mechanical Hamiltonian depend on atomic coordinates and therefore fluctuate in time.
However, we neglect the time dependence of the terms not explicitly expressed as a function of $\mathbf{q}^{N}_{i}$ in Eqs.~\ref{eq:ham}--\ref{eq:ham_vib} ($E_{S_{1}}-\Delta_{0-0}$, $\lambda$, $\lambda_{e}$, $\lambda_{h}$, $\omega_\mathrm{vib}$) under the assumption that their fluctuations are small and the spectroscopy can be adequately described using average values.

While many authors have included nonlocal electron-phonon couplings in tight-binding Hamiltonians like Eq.~\ref{eq:ham},\cite{coropceanu_charge_2007,lee_vibronic_2017,fetherolf_unification_2020,duan_ultrafast_2019} our approach is unique because it is able to handle nonlocal electron-phonon couplings of arbitrary form.
Typically, these couplings are treated in the linear response regime, assuming that lattice distortions are small enough that the electronic Hamiltonian is perturbed linearly.
Troisi and coworkers have also computed electron-phonon couplings without any assumptions of linearity,\cite{troisi_electronic_2005,troisi_dynamics_2006,arago_dynamics_2015,arago_regimes_2016,wang_multiscale_2010} but to our knowledge these calculations have not yet been incorporated into tight-binding approaches, due to computational limitations.
In organic crystals, nonlinear effects in nonlocal electron-phonon couplings may be important due to the sensitivity of the CT couplings to small geometric displacements.
The mixed quantum-classical method used here includes nonlocal electron-phonon coupling in a tight-binding Hamiltonian explicitly without the usual assumption of linearity.

To calculate the absorption spectrum, we represent the time-dependent Hamiltonian $\hat H_\mathrm{qm}(t)$ using a two-particle basis set,\cite{philpott_theory_1971} truncated to include only states with less than $v_\mathrm{max}$ vibrational quanta.
The maximum electron-hole separation is not restricted, except by the size of the simulated supercell.
We then compute the polarized absorption spectrum using the Fourier transform of the transition-dipole autocorrelation function $C_\mu(t)$
\begin{equation}
    \label{eq:absorption}
    \begin{split}
    A(\omega)&\propto \omega~\mathrm{Re} \int_{0}^{\infty} dt~e^{-i\omega t} C_\mu(t)\\
    C_\mu(t)&=\left\langle \boldsymbol\varepsilon\boldsymbol\cdot\boldsymbol\mu(t+t_0)
     \exp_+ \left\{-\frac{i}{\hbar}\int_{t_{0}}^{t} \, d\tau\,\hat{H}_\mathrm{qm}(\tau+t_{0})-\hat{\Gamma}(\tau+t_{0})\right\}\boldsymbol\varepsilon\boldsymbol\cdot\boldsymbol\mu\left(t_0\right)\right\rangle_{t_0}.
     \end{split}
\end{equation}
Here $\exp_+$ is a time-ordered exponential, $\boldsymbol\mu(t)$ contains the $\mathrm{S}_{1}\leftarrow \mathrm{S}_{0}$ transition dipole moments of the basis states at time $t$, $\boldsymbol\varepsilon$ is the polarization vector of the incoming light, $\hat{\Gamma}\left(t\right)$ is a matrix of eigenstate-dependent line broadening parameters that depends on the number of vibrational quanta in each eigenstate of $H_\mathrm{qm}(t)$ (see SM),\cite{yamagata_hj_2014} and $\left\langle\ldots\right\rangle_{t_0}$ is an average over initial time points of the MD trajectory.
Each spectrum is averaged over $N_\mathrm{samp}$ initial time points $t_0$ separated by $T_\mathrm{samp}$~(see SM).
The time-dependence of the transition dipole moments arises due to fluctuations in the molecular orientations according to $H_\mathrm{cls}$.
We solve Eq.~\ref{eq:absorption} using the Numerical Integration of the Schrodinger Equation approach.\cite{jansen_nonadiabatic_2006,jansen_waiting_2009}

\subsection{Molecular Dynamics Simulations\label{sec:md}}
We perform MD simulations of TAT and pentacene using the LAMMPS, VMD, and TopoTools packages.\cite{plimpton_fast_1995,humphrey_vmd_1996,kohlmeyer_topotools_2017}
We model the intra- and intermolecular potentials using the DREIDING force field\cite{mayo_dreiding_1990} with atomic charges taken from electronic structure calculations~(see SM).\cite{della_valle_computed_2008,strong_tetracene_2015,deng_predictions_2004}
This force field reproduces experimental crystal structures in similar systems.\cite{mattheus_modeling_2003}

The MD simulations are performed at constant number of particles, volume, and energy~(NVE) and use a rRESPA multi-timescale integrator\cite{tuckerman_reversible_1992} to accommodate the fast intramolecular motions~(see SM).
The TAT simulations are initialized in the crystal structure of Fan et al.\cite{fan_synthesis_2012} and the pentacene simulations are initialized in the crystal structure of Holmes et al.\cite{holmes_nature_1999}
We simulate a 10$\times$2$\times$2 supercell of TAT~(80 molecules) and a 4$\times$4$\times$2 supercell of pentacene~(64 molecules).
The dimensions of the supercells are chosen to be at least as large as twice the cutoff for the short-range interactions in the MD simulation and large enough to avoid finite size effects in the absorption spectra.
Each simulation is equilibrated for 10~ps during which velocities are rescaled every 1~ps to maintain a temperature of 298~K.
To compute the distribution of intermolecular geometries and couplings, we average over 1~ns of simulation time, collecting data every 1~ps.
To compute the spectra, we evaluate the Hamiltonian every 2~fs for 100~ps.

\section{Results and Discussion}
While the time-dependent quantities in equations~\ref{eq:ham}--\ref{eq:ham_vib} may be calculated directly from the atomic coordinates derived from MD simulations using standard ab-initio approaches,\cite{troisi_dynamics_2006,arago_dynamics_2015} modeling the spectroscopy of organic crystals in this way would require millions of such calculations because they must be repeated for each molecule, or pair of molecules, at every time step.
These requirements make such calculations nearly prohibitive, and at least impractical for many applications.
To render this approach practical and generalizable, we develop a map to estimate the values of the time-dependent quantities in the Hamiltonian~($E_\mathrm{CT}$, $J$, $t_e$, and $t_h$) for any realistic set of atomic coordinates, without the need for repeated electronic structure calculations. 
As discussed previously, this approach is motivated by similar methods that are used to model the infrared spectroscopy of condensed phases,\cite{baiz_vibrational_2020} and represents one of the main contributions of the current work.

For the CT energies $E_\mathrm{CT}$ and the Coulomb couplings $J$, relatively simple relationships between the atomic coordinates and the electronic properties have already been developed.
The CT couplings, on the other hand, are complex functions of the overlap between the frontier molecular orbitals of both chromophores involved, which are often oscillatory structures with many nodal surfaces.\cite{coropceanu_charge_2007,kazmaier_theoretical_1994,hestand_expanded_2018} 
While a similar mapping approach was recently used to account for diagonal disorder in organic semiconductors,\cite{shi_modeling_2018} the concept has not, to our knowledge, been applied to the CT couplings.
The main challenge of the problem is its dimensionality:
assuming that the coupling between two molecules is independent of the surrounding molecules, the map still must be parameterized in a $3\cdot2N-6\sim200$ dimensional space~($N=42$ for TAT and $N=36$ for pentacene).
To overcome this problem, we make the rigid-body approximation\cite{day_atomistic_2003,coropceanu_charge_2007} 
when evaluating $\hat{H}_\mathrm{qm}$.
Specifically, before computing the time-dependent quantities in the Hamiltonian~(Eq.~\ref{eq:ham}), we replace each molecule in the MD simulations with the geometry-optimized monomer structure, translated and rotated to have the same center of mass and principal axes of inertia.
In some respects, it would be simpler to perform an MD simulation with rigid molecules.
Instead, we use the DREIDING force field, which calls for flexible molecules, because it is well tested for organic crystals like these and is easily transferable to new systems without extensive parameterization efforts.\cite{mayo_dreiding_1990,strong_tetracene_2015,mattheus_modeling_2003}
Simulating rigid monomers would require reparameterizing the MD force field in this work, as well as in future studies of different organic crystals.

The rigid-body approximation decouples the intra- and intermolecular phonons such that the nonlocal electron-phonon coupling depends only on the intermolecular modes and contributions from the intramolecular modes are neglected.
Note, however, that the important high-frequency intramolecular modes are still treated through the local electron-phonon coupling~(Eq.~\ref{eq:ham_vib}).
Importantly, the rigid-body approximation reduces the dimensionality of the problem from $\sim$200 to 6. 
Since each monomer is exactly the same, the relative coordinates of any pair can be specified by 3 translational and 3 rotational degrees of freedom.
In this reduced dimension, it might be possible to perform an explicit interpolation on a 6-dimensional grid that maps nearest-neighbor intermolecular geometries observed in the simulations to precomputed couplings at sparse grid points.
We first test this idea with TAT because it is simpler than pentacene in regard to modeling optical properties; TAT can be modeled as a collection of non-interacting one-dimensional $\pi$-stacks\cite{yamagata_hj_2014} while pentacene must be treated as a collection of two-dimensional layers~(Fig.~\ref{fig:schematic}).\cite{hestand_exciton_2015}

\subsection{TAT}
\subsubsection{Developing the Map\label{sec:TATmap}}
\begin{figure}
    \centering
    \includegraphics{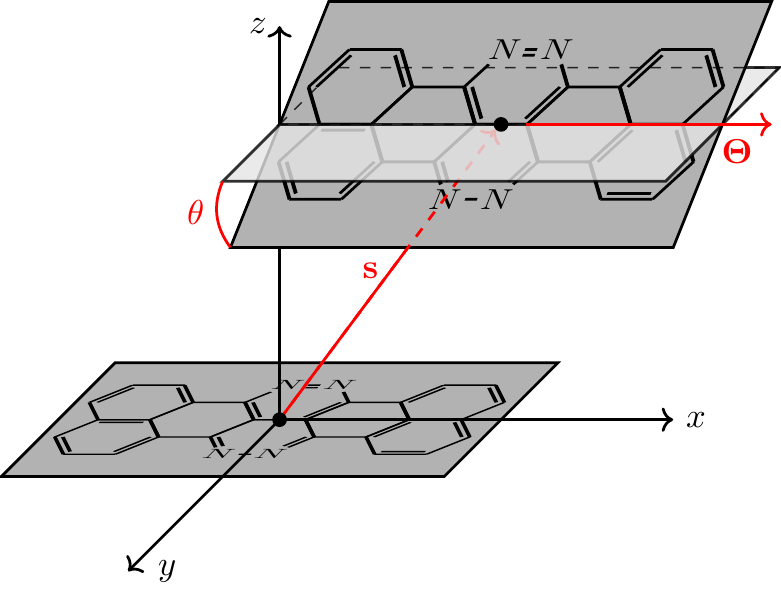}
    \caption{Coordinate system used to define the relative orientation of two molecules in a dimer pair.
    Here, the reference TAT molecule is centered at the origin with its inertial axes defining the $x$, $y$, and $z$ axes.
    The second TAT molecule is translated by the vector $\mathbf{s}$, then rotated about the axis $\mathbf{\Theta}$ by the angle $\theta$.\label{fig:coordinateSystem}}
\end{figure}

To assess the feasibility of an interpolative map, we quantify the fluctuations of the 6 intermolecular degrees of freedom for all nearest-neighbor pairs of molecules in TAT in terms of the quantities $\mathbf{s}$, $\theta$, and $\mathbf{\Theta}$~(Fig.~\ref{fig:coordinateSystem}).
\begin{figure}[ht!]
\includegraphics{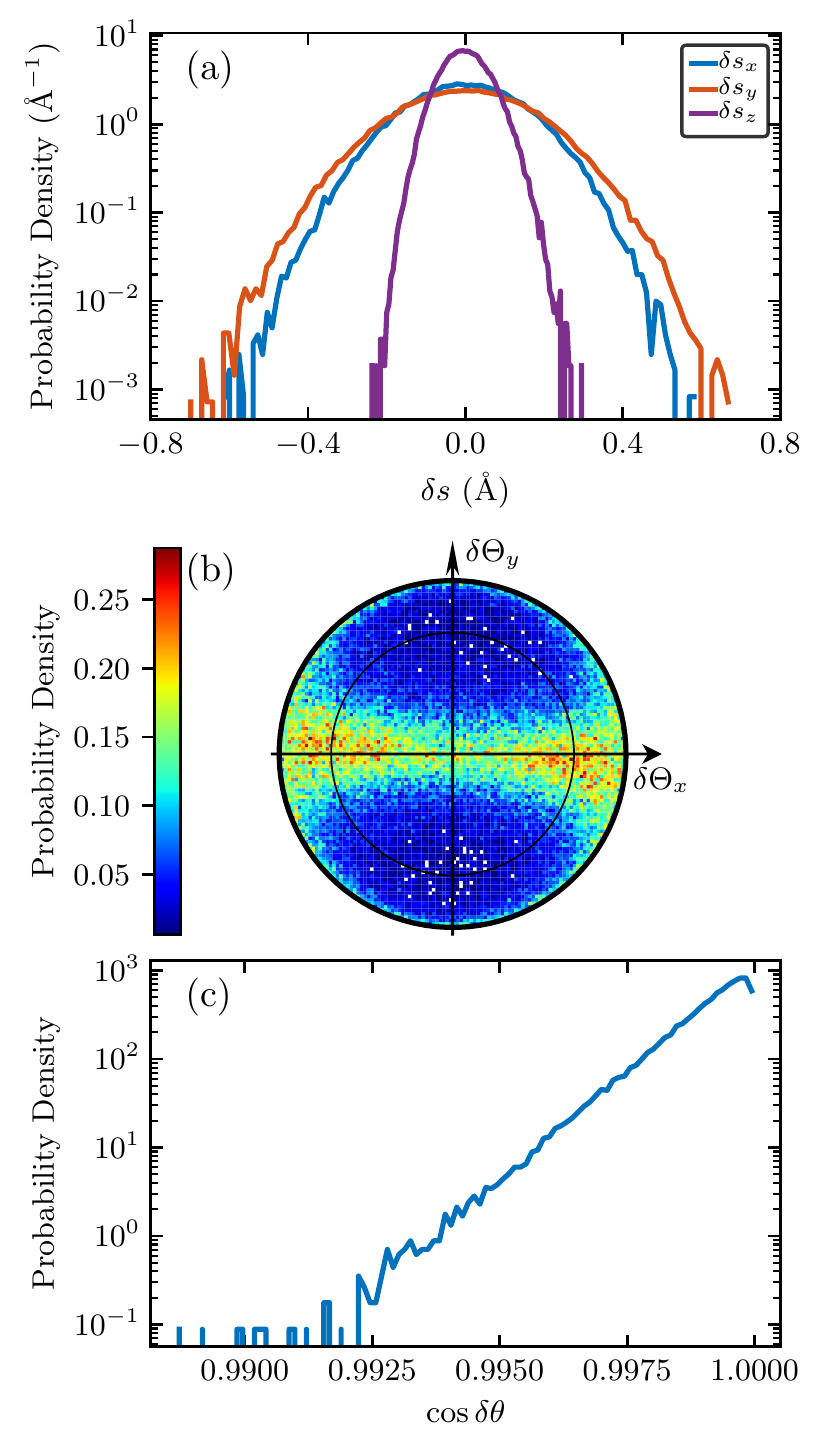}
\caption{Distributions of nearest-neighbor pair geometries in TAT.
The coordinate system is defined in Figs.~\ref{fig:schematic}a and \ref{fig:coordinateSystem}.
(a) Distributions of the fluctuations of the translational slips. 
(b) Distribution of the fluctuations of the rotational axes on a unit sphere, depicted as a Lambert azimuthal equal-area projection in which the $\delta\Theta_z=1$ pole of the unit sphere is mapped to the origin, the $\delta\Theta_z=0$ equator is shown as the thin black line, and the $\delta\Theta_z=-1$ pole is mapped to the thick black circle at the perimeter.
This projection conserves area, which is an important property for a histogram.
(c) Distribution of the cosine of the fluctuations of the rotational angles. 
It is important to plot the cosine instead of the angle itself to avoid singularities in the Jacobian at $\delta\theta=0^\circ$.\cite{strong_tetracene_2015}}
\label{fig:TATdimers}
\end{figure}
The 3 translational degrees of freedom are described as ``slips'' $\mathbf{s}_{ij}$ of molecule $j$'s center-of-mass along the principal axes of molecule $i$.
The 3 orientational degrees of freedom are described by an angle of rotation $\theta_{ij}$~(1 degree of freedom) and the rotation axis $\mathbf\Theta_{ij}$ about which to rotate~(3 degrees of freedom minus 1 for normalization).

We are interested in the fluctuations of the intermolecular geometries about their averages, which we define as $\delta\mathbf{s}$, $\delta\theta$, and $\delta\mathbf\Theta$.
For the translational slips, $\delta\mathbf{s}$ is simply defined by $\delta\mathbf{s} = \mathbf{s} - \langle \mathbf{s}\rangle$.
The definitions of $\delta\theta$ and $\delta\mathbf\Theta$ are more complex, and are discussed in the SM.
In the case of TAT, the molecules are $\pi$-stacked, so their equilibrium structure has no rotation ($\langle\theta\rangle=0^\circ$, Table~\ref{tab:TATgrid}).
In this special case, $\theta=\delta\theta$ and $\mathbf\Theta=\delta\mathbf{\Theta}$ so these distinctions are unimportant, but as we will see, this is not the case for pentacene~(Sec.~\ref{sec:pen}).
Several other details regarding the intermolecular degrees of freedom are discussed in the SM.

We find that the distributions of slips and orientations are localized~(Fig.~\ref{fig:TATdimers}), making this system amenable to an interpolation map as described above.
It is not surprising that the slip distributions and rotation angle distribution are localized, since the system is crystalline so mobility is limited, but it was not clear to us \textit{a priori} that the distribution of rotation axes would likewise be localized.
Specifically, we find that the observed rotation axes $\delta\mathbf{\Theta}$ cluster around the $xz$-equator of the unit sphere.
This important observation means that instead of covering the entire surface of the unit sphere with an interpolation grid of rotational axes, we only need to include rotational axes along the $xz$-equator.

Based on the distributions in Fig.~\ref{fig:TATdimers}, we construct a grid that encompasses most of the observed configurations~(Table~\ref{tab:TATgrid}).
\begin{table}[t]
\centering
\caption{The interpolation grid used to calculate the CT couplings for TAT.
The origin of the grid is at the experimental geometry~(see text).
The grid includes all points that are integer multiples of the grid spacing and no more than a distance of ``extent'' from the origin.
The rotation axis $\mathbf\Theta$ grid starts at $(1,0,0)$, and includes the 6 axes that are equally spaced at 60$^\circ$ intervals about the $xz$-equator.
}
\label{tab:TATgrid}
\begin{tabular}{l S[table-format=1.2] S[table-format=1.2] S[table-format=2.1] S[table-format=1.1] S[table-format=1]}
\hline
              & {Expt.\cite{fan_synthesis_2012}} & {$\langle \cdot \rangle_\mathrm{sim}$} & {Extent} & {Spacing} & {Grid Size}   \\
\hline
$s_x$ (\AA{})            & 1.00 & 1.01  & 0.6   & 0.2       & 7  \\
$s_y$ (\AA{})            & 1.28 & 1.12  & 0.6   & 0.2       & 7  \\
$s_z$ (\AA{})            & 3.37 & 3.42  & 0.2   & 0.2       & 3  \\
$\theta$ ($^{\circ}$)    & 0    & 0     & 10    & 5         & 3  \\
$\mathbf\Theta$                     & {--} & {--}  & {$xz$-equator}  & {60$^\circ$} & 6  \\
\hline
\end{tabular}
\end{table}
The sparsity of the grid is informed by calculations of CT couplings for TAT and similar $\pi$-conjugated systems,\cite{hestand_interference_2015,hestand_exciton_2015} which show that the couplings vary approximately linearly on the scale of the grid spacings we use.
This grid results in a total of 2646 grid points, or 2646 electronic structure calculations of dimers that must be completed to compute the CT couplings at each grid point.
This number can be reduced to 1911 points by recognizing that for grid points with $\delta\theta=0$ the rotational axis is irrelevant and that dimension of the grid can be ignored.
The size of the grid could be further reduced by accounting for the symmetry of the molecule and the cross-correlation between the distributions of intermolecular geometries.
That is, all the slips do not take their maximum values at the same time.
For the level of electronic structure calculations we use, and for the number of atoms in TAT, these optimizations are not necessary so we do not make them.
They may become necessary for larger molecules or more expensive electronic structure basis sets.

The CT couplings are calculated at each grid point of the map following the methods discussed in Refs.~\citenum{senthilkumar_charge_2003,senthilkumar_absolute_2005,valeev_electronic_2006,mikolajczyk_long-range_2011} using the Gaussian software package\cite{frisch_gaussian_2010} and the B3LYP/3-21G level of theory.
The map is provided as a supplementary data file~(see SM).
Several other basis sets and functionals were also considered, but we found that all methods gave qualitatively similar results~(see SM).
The couplings for non-nearest-neighbors are assumed to be zero, which is generally a good approximation given the short-range, exponentially decaying nature of these interactions.\cite{coropceanu_charge_2007}

The average nearest-neighbor pair geometry we observe in simulation is quite close to the experimental crystal structure; the largest difference in slip is 0.16~\AA{} ~(Table~\ref{tab:TATgrid}).\cite{fan_synthesis_2012}
Even over such small displacements, however, the computed CT couplings can change sign.\cite{hestand_exciton_2015}
To account for this sensitivity, we treat the fluctuations about the equilibrium geometry in the MD simulation as fluctuations about the experimental crystal structure. 
That is, in the MD simulation we measure the slips $\mathbf{s}=\langle \mathbf{s}\rangle + \delta \mathbf{s}$, but we interpolate the CT couplings using $\mathbf{s}=\mathbf{s}_\mathrm{expt} + \delta \mathbf{s}$.
This provides the important benefit that the interpolation map only needs to be computed once, and can then be applied to any MD simulation, regardless of the equilibrium intermolecular geometries that are realized by a particular force field.
At most, one might have to extend the map to accommodate larger fluctuations in one MD simulation relative to another.

With the completed grid of CT couplings, we perform a linear interpolation to map a dimer configuration from the MD simulation to a pair of CT couplings $t_e$ and $t_h$.
We interpolate the orientational axis by first projecting the axis to the $xz$-equator and then interpolating along the equator.
For the 0.1\% of dimer configurations that are outside the grid, we linearly extrapolate the couplings.

We now turn to the calculation of the time-dependent CT energies~($E_\mathrm{CT}$) and Coulomb couplings~($J$).
To compute the CT energies, we treat each electron and hole as a point charge located at the center-of-mass of its chromophore.
This is a good approximation for large electron-hole separations, and has been applied successfully to compute time-independent CT energies in previous work.\cite{hestand_expanded_2018,yamagata_hj_2014,hestand_polarized_2015}
Moreover, it allows us to replace the expression for the Coulomb  potential $V\left(\mathbf{q}^{N}_i(t),\mathbf{q}^{N}_j(t)\right)$ in Eq.~\ref{eq:ect} with a simple expression
\begin{equation}\label{eq:ct_scaling}
    V\left(r_{ij}^\mathrm{com}(t)\right)=\frac{-e^2}{4\pi \varepsilon_{0}\varepsilon_{s}}\frac{1}{r_{ij}^\mathrm{com}(t)}.
\end{equation}
Here $r_{ij}^\mathrm{com}(t)=|\mathbf{q}_j^{\mathrm{com}}(t) - \mathbf{q}_i^{\mathrm{com}}(t)|$ is the distance between the center of mass coordinates of molecules $i$ and $j$, $\varepsilon_s$ is the static dielectric constant, and the remaining terms take their usual meanings.
The CT energy therefore fluctuates with the center of mass fluctuations, $E_\mathrm{CT}\left(r_{ij}^\mathrm{com}(t)\right)$.
It is convenient to express $E_\mathrm{CT}\left(r_{ij}^\mathrm{com}(t)\right)$ in terms of the static nearest-neighbor CT energy  $E_\mathrm{CT}(\langle r_{n.n.} \rangle)=I_{P}-E_{A}+P+V\left(\langle r_{n.n.} \rangle \right)$, so that
\begin{equation}\label{eq:ct_neighbor}
    E_\mathrm{CT}\left(r_{ij}(t)\right)=E_\mathrm{CT}(\langle r_{n.n.} \rangle)-V\left(\langle r_{n.n.}\rangle \right)+V\left(r_{ij}^\mathrm{com}(t)\right).
\end{equation}
While $E_\mathrm{CT}(\langle r_{n.n.}\rangle)$ could be computed from first principles, previous work has either treated it as a fitting parameter, whose value is extracted by fitting calculated spectra to experimental spectra,\cite{yamagata_hj_2014, hestand_exciton_2015} or derived it from experiment.\cite{yamagata_nature_2011,hestand_polarized_2015}
We take the nearest-neighbor CT energy for TAT from Yamagata et al.~(see SM).\cite{yamagata_hj_2014}

The Coulomb coupling may be calculated using a variety of approaches, including the point dipole approximation, transition charges,\cite{chang_monopole_1977} and the density-cube method.\cite{Krueger1998}
Here we use the transition charge method as it provides a good compromise between accuracy and speed.\cite{kistler_benchmark_2013}
Because the transition charges depend on the intramolecular geometry and the intramolecular geometry of each molecule fluctuates during the MD simulations, computing the Coulombic coupling for these structures would necessitate an expensive calculation of the transition charges of each molecule in the crystal at each time step.
As discussed in the context of the CT couplings above, we make the rigid-body approximation and replace each molecule in the MD simulation with a geometry-optimized molecule.
This means that we need only compute the transition charges once, for the geometry-optimized monomer~(see SM).
Within this framework, it is straightforward to apply the transition charges of the geometry-optimized molecule at each time step and compute the Coulomb coupling between each pair.
The raw Coulombic couplings are then screened by an optical dielectric constant $\varepsilon$.

We note that one must ensure that the phases of the CT couplings and transition charges are correct and consistent for the duration of the simulation.
We assign the phases by visual inspection and ensure that they are consistent with previous works.\cite{yamagata_hj_2014}
A complete description of the relative phases necessary for using the maps provided in the supplementary data files is provided in the SM.

The values of all time-independent parameters are given in the SM.

\subsubsection{Spectroscopy}
\begin{figure}[t!]
\includegraphics{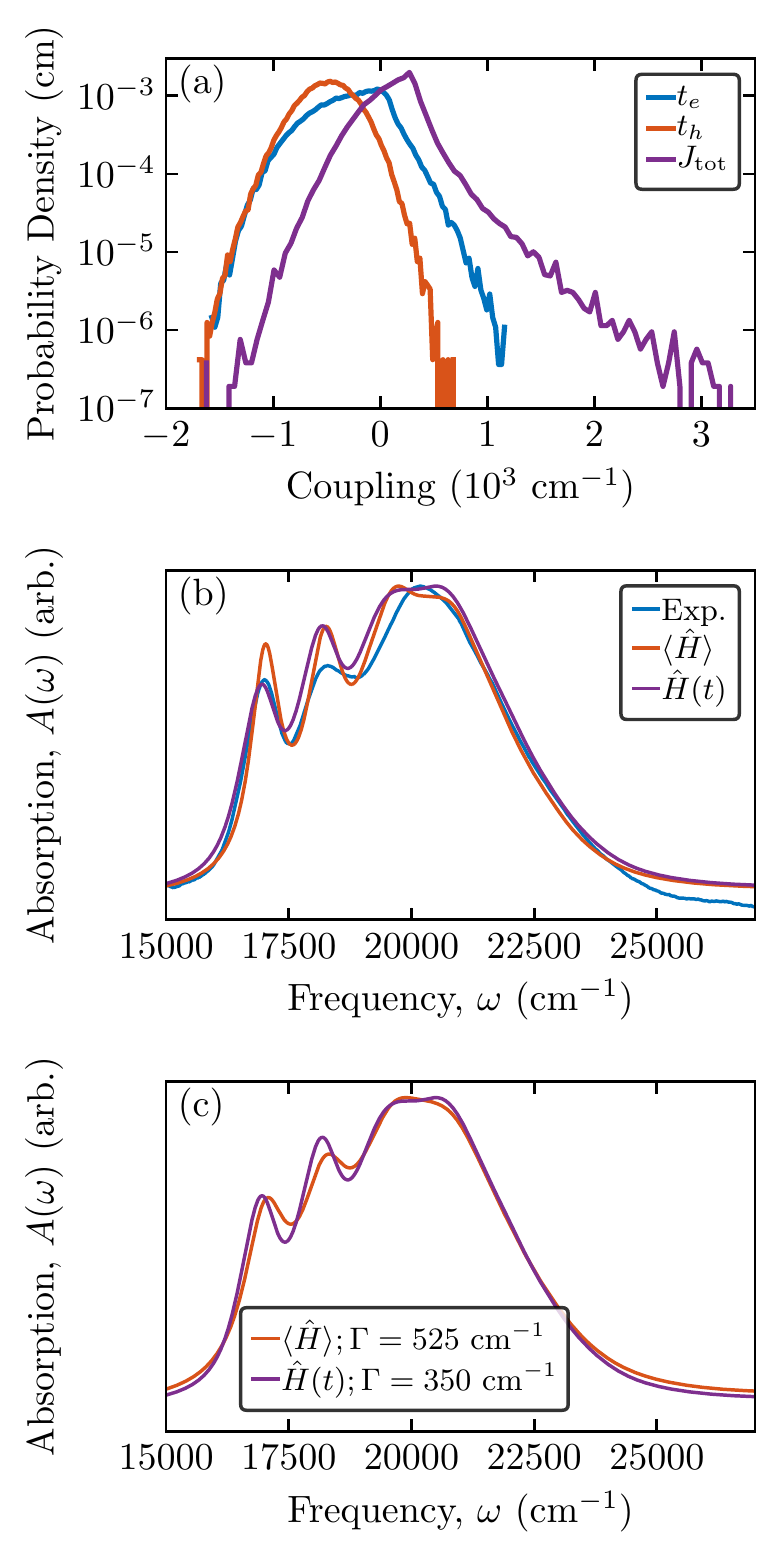}
\caption{(a) The distributions of the nearest-neighbor couplings in TAT. (b) The experimental absorption spectrum of TAT~(blue),\cite{yamagata_hj_2014} compared to the theoretical spectra with~(purple) and without~(orange) nonlocal electron-phonon coupling.
The spectra are normalized by the maximum peak height to allow comparison between the experiment and the theory.
(c) The effects of the nonlocal electron-phonon coupling cannot be fully captured by simply broadening the $\langle\hat{H}\rangle$ spectrum, even though the change in the 0-0/0-1 peak area ratio can~(see SM).
The purple curve here is the same as the purple curve in panel (b).}
\label{fig:TATspec}
\end{figure}

We first compute the distributions of the couplings in TAT.
We find that the fluctuations in the couplings are on the same order of magnitude as their means, in line with observations of Troisi and coworkers~(Fig.~\ref{fig:TATspec}a, SM).\cite{troisi_dynamics_2006,arago_dynamics_2015}
To make contact with their work on fluctuations in the exciton coupling, we also compute the total exciton coupling $J_\mathrm{tot} = J_\mathrm{SR}+J$, where $J$ is the Coulomb coupling in Eq.~\ref{eq:ham_fe} and $J_\mathrm{SR}$ is the short-ranged, charge-transfer mediated coupling\cite{harcourt_rate_1994,scholes_rate_1995}
\begin{equation}
    J_\mathrm{SR}(\mathbf{q}^{N}_i,\mathbf{q}^{N}_j) = \frac{-2t_e(\mathbf{q}^{N}_i,\mathbf{q}^{N}_j) t_h(\mathbf{q}^{N}_i,\mathbf{q}^{N}_j)}{E_\mathrm{CT}(\mathbf{q}^{N}_i,\mathbf{q}^{N}_j) - E_{S_1} - \Delta_{0-0}}.
\end{equation}
Fig.~\ref{fig:TATspec}a shows the distribution of $J_\mathrm{tot}$ for nearest-neighbors in TAT.
Note that $J_\mathrm{SR}$ and $J_\mathrm{tot}$ do not enter into the calculation of the spectrum, but are simply used to make a comparison with previous results.
As discussed in Sec.~\ref{sec:intro}, one might expect these large fluctuations to have important effects on the absorption spectroscopy of TAT, but this expectation is at odds with the successful modeling of the absorption of TAT without nonlocal electron-phonon coupling.\cite{yamagata_hj_2014}
Further, the coupling distributions are strongly non-Gaussian, indicating that nonlinear models of nonlocal electron-phonon coupling, like the one presented here, may be needed to accurately capture photophysical and charge transport properties.
A direct comparison between linear and nonlinear models of nonlocal electron-phonon coupling is beyond the scope of this work, but merits future investigation.

The effects of these fluctuating couplings on the theoretical absorption spectrum of TAT are shown in Fig.~\ref{fig:TATspec}b.
To isolate the effects of nonlocal electron-phonon coupling, we compare the spectrum computed with the time-averaged Hamiltonian $\langle \hat H \rangle$ to the time-dependent Hamiltonian $\hat H(t)$~(Eq.~\ref{eq:ham}).
The spectrum is calculated for only one of the eight pillars of molecules in the MD simulation~(Fig.~\ref{fig:schematic}c), because the couplings between adjacent stacks are negligible, so the stacks can be treated as independent, quasi-one-dimensional systems.\cite{yamagata_hj_2014}

The theoretical spectra~(Fig.~\ref{fig:TATspec}b), both with and without nonlocal electron-phonon coupling, match the experimental spectrum quite well, as has been previously demonstrated for the case without nonlocal electron-phonon coupling.\cite{yamagata_hj_2014}
Note that the $\langle \hat{H}(t)\rangle$ spectrum is calculated using the same parameter set and model as in Ref.~\citenum{yamagata_hj_2014}, with only minor changes in the couplings.
The nonlocal electron-phonon coupling has two noticeable effects on the spectrum: the spectrum broadens and the relative peak heights change.
To quantify these changes, we fit the spectra with a set of four Lorentzian functions, one for each of the four observable vibronic peaks~(see SM).
We find that the full-width half-maximum of the vibronic peaks increase by an average of~$\sim200$ \cm{} due to the nonlocal electron-phonon coupling, and that the change in the relative peak heights is mainly a byproduct of the broadening, not a separate effect~(see Fig.~\ref{fig:TATspec}c and SM).
Note, however, that the entire spectral line shape of the $\hat{H}(t)$ spectrum cannot be reproduced by uniformly broadening the $\langle \hat{H} \rangle$ spectrum~(Fig.~\ref{fig:TATspec}c), indicating that nonlocal electron-phonon coupling has a more nuanced effect on the spectrum.
This is discussed in detail in Sec.~\ref{sec:penSpec} when considering the pentacene spectrum.

These results provide proof-of-principle for our approach to incorporating nonlocal electron-phonon coupling into spectroscopic calculations.
Despite the large fluctuations in the CT couplings~(Fig.~\ref{fig:TATspec}a), the effects on the absorption spectrum are, somewhat surprisingly, relatively minimal.
In order to make contact with previous work studying the fluctuations of couplings in acene systems,\cite{troisi_dynamics_2006,arago_dynamics_2015,arago_regimes_2016,fornari_exciton_2016} and to demonstrate our mapping approach in a more complex crystal structure, we now turn to the case of pentacene.

\subsection{Pentacene\label{sec:pen}}
\subsubsection{Developing the Map}
We begin, as in the case of TAT, by quantifying the fluctuations in the nearest-neighbor intermolecular geometries observed in pentacene.
In TAT, only the nearest-neighbors, which form a one-dimensional stack of molecules along the crystalline $\mathbf{a}$-axis, have non-negligible couplings.
In pentacene, there are non-negligible couplings between neighbors in both the $\mathbf{a}$ and $\mathbf{b}$ directions, resulting in an effective two-dimensional system that must be considered to accurately model spectroscopy or charge transport.
This means that we can no longer consider only nearest-neighbors, but must consider all neighbors within the crystalline $\mathbf{ab}$-plane~(Fig.~\ref{fig:schematic}).
In pentacene, these neighbors can be classified by their spatial relationship in the crystal structure.
We label them using their fractional coordinates in the unit cell relative to a reference molecule at $(0,0,0)$.
In this notation, the two nearest-neighbors are those at $\pm(\half,-\half,0)$ and $\pm(\half,\half,0)$, with center-of-mass distances of 4.7 and 5.2~\AA{}, respectively.\cite{holmes_nature_1999}
These pairs exhibit the herringbone stacking motif that characterizes crystalline acenes.
The next nearest-neighbors are the $\pm(1,0,0)$ and $\pm(0,1,0)$ dimers at 6.3 and 7.7~\AA{}.\cite{holmes_nature_1999}
These pairs are co-planar, not herringbone stacked.
The CT couplings ($t_e$ and $t_h$) between the $\pm(0,1,0)$ dimers are negligible,\cite{yamagata_nature_2011,hestand_polarized_2015} so we ignore them here, but this still leaves three distinct types of dimers, compared to a single type in TAT.
This means that we must parameterize the interpolation map in three disjoint regions of the 6-dimensional intermolecular geometry space.
Note that the labelling scheme described above is the same as that of Hestand et al.\cite{hestand_polarized_2015}, except that we define the lattice vectors in the same way as Holmes et al.,\cite{holmes_nature_1999} while Hestand et al. swap the $\mathbf{a}$ and $\mathbf{b}$ lattice vectors.

\begin{figure}[ht!]
\includegraphics{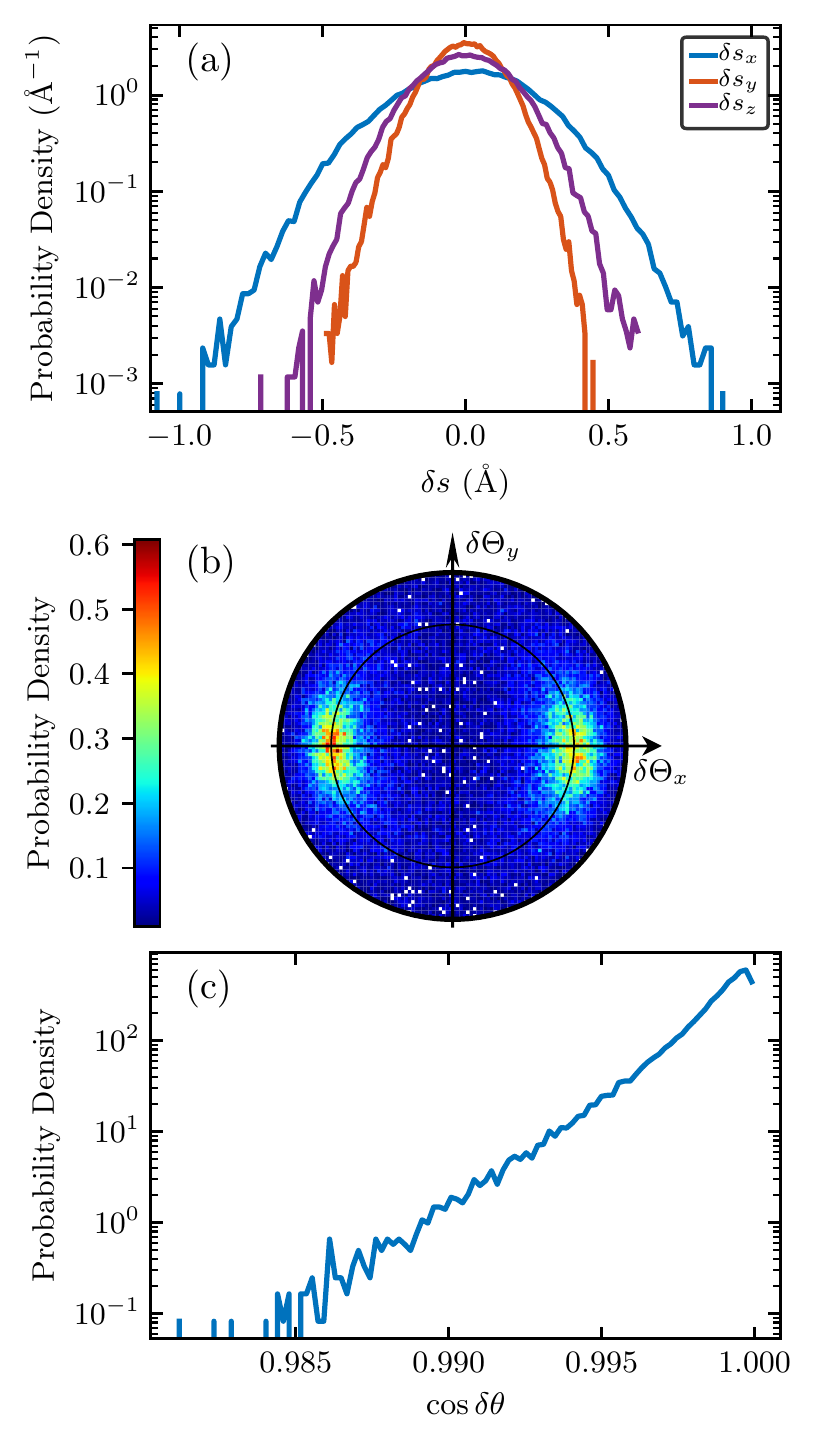}
\caption{Distributions of the geometries of the $\pm(\half,-\half,0)$ dimer in pentacene.
The distributions for the $\pm(\half,\half,0)$ and $\pm(1,0,0)$ dimers are shown in the SM.
The molecular coordinate system is defined in Fig.~\ref{fig:schematic}d.
(a) Distributions of the fluctuations of the translational slips. 
(b) Distribution of the fluctuations of the rotational axes on a unit sphere, depicted as a Lambert azimuthal equal-area projection.
(c) Distribution of the cosine of the fluctuations of the rotational angles.
See the caption of Fig.~\ref{fig:TATdimers} for details.}
\label{fig:PENTdimers}
\end{figure}

The distributions of intermolecular geometries for the $\pm(\half,-\half,0)$ dimer are shown in Fig.~\ref{fig:PENTdimers}.
The distributions look qualitatively similar to those in TAT with one exception:
the fluctuations in the rotational axis are localized around $(\pm 1,0,0)$.
This means that most rotations are about the molecular long axis of pentacene.
The distributions for the $\pm(\half,\half,0)$ and $\pm(1,0,0)$ dimers are qualitatively similar~(see SM).
Based on these distributions, we construct a grid for each type of dimer, similar to the one for TAT shown in Table~\ref{tab:TATgrid}~(see SM).
Instead of interpolating the rotational axes along the $xz$-equator, we interpolate all rotational axes with $\delta\Theta_x\leq0$ to $(-1,0,0)$ and all rotational axes $\delta\Theta_x>0$ to $(1,0,0)$.
The total number of grid points is 11,277, much larger than for TAT due to the three disjoint regions of the map that must be considered.
As before, we shift the origin of the grid to the experimental dimer configurations, making the grid generalizable to any MD force field.
Previous work on pentacene has scaled the CT couplings obtained from DFT calculations by a factor of 1.1 to obtain better agreement with experiment.\cite{hestand_polarized_2015}
Here we do not scale the CT couplings to be consistent with the calculations for TAT.

As before, the fluctuating Coulomb couplings are computed based on the transition charge method, with transition charges calculated for the geometry-optimized monomer~(see SM).
The CT energies are also computed as before~(Eq.~\ref{eq:ct_neighbor}), except that the nearest-neighbor CT energy is parameterized separately for each dimer type~(see SM), in accordance with previous theoretical\cite{yamagata_nature_2011,beljonne_charge_2013,hestand_polarized_2015} and experimental works.\cite{sebastian_charge_1981}
For all electron-hole pairs further than the $\pm(0,1,0)$ dimer, the CT energy is calculated using the $\pm(\half,-\half,0)$ energy and Eq.~\ref{eq:ct_neighbor}.\cite{yamagata_nature_2011,beljonne_charge_2013,hestand_polarized_2015,hestand_polarized_note_axes}

\subsubsection{Spectroscopy\label{sec:penSpec}}
\begin{figure}[ht!]
\includegraphics{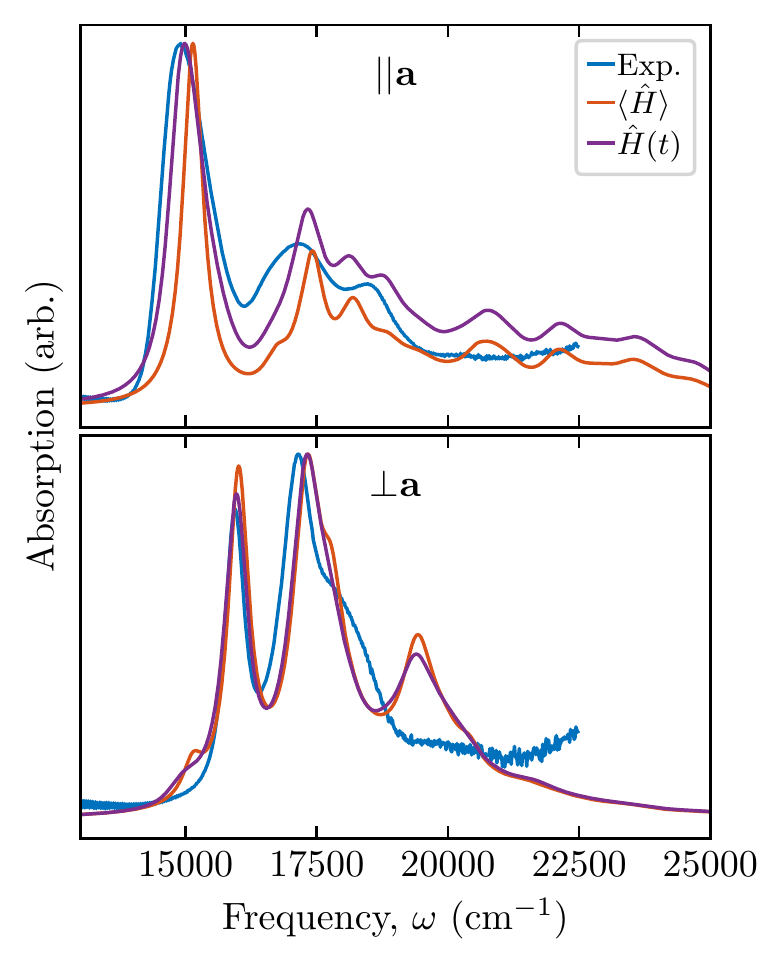}
\caption{The experimental absorption spectrum of pentacene, polarized along the crystalline $\mathbf{a}$-axis~(top) and perpendicular to the $\mathbf{a}$-axis~(bottom).
We compare the experimental spectra~(blue), to the theoretical spectra with~(purple) and without~(orange) nonlocal electron-phonon coupling.
The spectra are normalized to the maximum peak height to allow comparison between the experiment and the theory.}
\label{fig:PENTspec}
\end{figure}

The calculated $||\mathbf{a}$ and ${\perp}\mathbf{a}$ polarized spectra with and without nonlocal electron-phonon couplings are shown in Fig.~\ref{fig:PENTspec}.
Both theoretical spectra agree qualitatively with the experiment.\cite{hestand_polarized_2015,hestand_polarized_note_axes}
As in the case of TAT, the main effect of nonlocal electron-phonon coupling is to broaden the spectra.
Interestingly, however, the $||\mathbf{a}$ spectrum~(lower Davydov component) broadens considerably more than the ${\perp} \mathbf{a}$ spectrum~(upper Davydov component). 
Fitting the spectra with Lorentzian functions shows that the full-width half-maximum of the lowest energy vibronic peak increases by about 300 \cm{} in the $||\mathbf{a}$ spectrum when nonlocal electron-phonon coupling is included, but only by about 50~\cm{} in the ${\perp}\mathbf{a}$ spectrum~(see SM).
Previous theoretical modeling of the absorption spectrum of pentacene has used a larger line width parameter for the $||\mathbf{a}$ spectrum than the ${\perp}\mathbf{a}$ spectrum in order to capture the observed experimental line widths.\cite{hestand_polarized_2015,a_hestand_polarized_note}
Here, we use the same line width parameter for both components~(see SM) and find that the nonlocal electron-phonon coupling selectively increases the line width of the $||\mathbf{a}$ spectrum.

The disparate line width broadening of the two polarization components can be understood by considering a simple dimer model that incorporates nonlocal electron-phonon coupling through fluctuations in $t_{e}(t)$ and $t_{h}(t)$.
To focus on the effect of nonlocal electron-phonon coupling, we neglect local vibronic coupling and Coulomb coupling.
The Hamiltonian for this simplified system can be expressed in the basis $\left|h_{1}, e_{1} \right\rangle$, $\left|h_{2}, e_{2} \right\rangle$, $\left|h_{1}, e_{2} \right\rangle$, $\left|h_{2}, e_{1} \right\rangle$ as
\begin{equation}
    \hat{H}_\mathrm{dimer}(t) = \begin{bmatrix}
     E_\mathrm{S_{1}}        & 0                       & t_{e}(t) & t_{h}(t) \\
     0                       & E_\mathrm{S_{1}}        & t_{h}(t) & t_{e}(t) \\
     t_{e}(t) & t_{h}(t) & E_\mathrm{CT}           & 0 \\
     t_{h}(t) & t_{e}(t) & 0                       & E_\mathrm{CT} \\
    \end{bmatrix}.
\end{equation}
Here $\left|h_{n},e_{m} \right\rangle$ represent a state with a hole on chromophore $n$ and an electron on molecule $m$.
When $n=m$, the state is a Frenkel exciton and when $n\ne m$ the state is a CT exciton.

It is convenient to transform the Hamiltonian to a symmeterized basis $\left|\psi_\mathrm{FE+}\right\rangle$, $\left|\psi_\mathrm{CT+}\right\rangle$, $\left|\psi_\mathrm{FE-}\right\rangle$, $\left|\psi_\mathrm{CT-}\right\rangle$ (see SM).
In this basis, only states with the same symmetry~($+$ or $-$) are coupled, and the Hamiltonian is
\begin{equation}
    \hat{H}_\mathrm{dimer}(t) = \begin{bmatrix}
     E_\mathrm{S_{1}} & t_{+}(t) & 0 & 0 \\
    t_{+}(t) & E_\mathrm{CT} & 0 & 0 \\
     0 & 0 & E_\mathrm{S_{1}} & t_{-}(t) \\
     0 & 0 & t_{-}(t)  & E_\mathrm{CT}\\
    \end{bmatrix}.
\end{equation}
where $t_\pm(t) = t_{e}(t) \pm t_{h}(t)$.
In this representation, the coupling between $\left|\psi_\mathrm{FE+}\right\rangle$ and $\left|\psi_\mathrm{CT+}\right\rangle$ depends on the sum of the CT couplings $t_+(t)=t_{e}(t)+t_{h}(t)$ while the coupling between $\left|\psi_\mathrm{FE-}\right\rangle$ and $\left |\psi_\mathrm{CT-}\right\rangle$ depends on the difference $t_-(t)=t_{e}(t)-t_{h}(t)$.
While this discrepancy is rather subtle, it has profound effects on how nonlocal electron-phonon coupling can influence the eigenstates of the system and the accompanying absorption spectrum.

Previous studies based on Frenkel exciton models have shown that off-diagonal disorder broadens the line width according to the variance of the disorder distribution; broader disorder distributions give rise to broader line widths.\cite{klugkist_scaling_2008,fidder_optical_1991}
In our model, then, the line width of transitions to the symmetric states are broadened by the breadth of the distribution of $t_{+}(t)$ while the line width of transitions to the antisymmetric states are broadened by the breadth of the distribution of $t_{-}(t)$.
To understand how disparate broadening between the two states might arise, consider the case where the fluctuations in $t_e(t)$ and $t_h(t)$ are Gaussian with means $\langle t_e\rangle$ and $\langle t_h\rangle$ and variances $\sigma_e^2$ and $\sigma_h^2$.\cite{akimov_stochastic_2017}
We need not specify anything about the temporal correlations of $t_e(t)$ and $t_h(t)$, except that they are shorter-lived than the timescale of the spectroscopic measurement, so that the Gaussian statistics are sufficiently sampled.
If the fluctuations in $t_{e}(t)$ and $t_{h}(t)$ are uncorrelated, then the variances of the distributions of $t_{+}(t)$ and $t_{-}(t)$ are both $\sigma_\pm^2 = \sigma_e^2+\sigma_h^2$.
In this case, the absorption peaks arising from the symmetric and antisymmetric states broaden to the same extent.
In real systems, however, $t_{e}(t)$ and $t_{h}(t)$ \textit{are} correlated, so $\sigma_\pm^2 = \sigma_e^2+\sigma_h^2\pm 2\rho\sigma_e^{}\sigma_h^{}$, where $\rho$ is the correlation coefficient between the distributions of $t_{e}(t)$ and $t_{h}(t)$.
Thus, when $\rho$ is positive, $\sigma_+^2>\sigma_-^2$, and vice versa for $\rho<0$.
This means that the absorption peaks arising from the symmetric states are broadened more~($\rho>0$) or less~($\rho<0$) by the nonlocal electron-phonon coupling than those arising from the antisymmetric states.
In the limiting case where $t_e(t)$ and $t_h(t)$ are perfectly correlated ($\rho=1$) and identically distributed~($\sigma_e=\sigma_h$),  $\sigma_{+}^2=4\sigma_e^2$ while $\sigma_-^2=0$.
In this extreme, the nonlocal electron-phonon coupling \textit{only} broadens the symmetric absorption peak.
Likewise when $t_e(t)$ and $t_h(t)$ are perfectly anticorrelated ($\rho=-1$) and identically distributed, only the antisymmetric absorption peak is broadened.

Thus, correlations between the fluctuations of $t_e(t)$ and $t_h(t)$ can lead to selective broadening like that seen in pentacene.\cite{hestand_polarized_2015,hestand_polarized_note_axes}
In pentacene, the $||\mathbf{a}$ polarized lower Davydov component arises due to absorption to the symmetric state while the ${\perp}\mathbf{a}$ polarized upper Davydov component spectrum arises mainly due to absorption to the antisymmetric state.\cite{yamagata_nature_2011, hestand_polarized_2015,hestand_polarized_note_axes}
Our simple model therefore demonstrates that the selective broadening of the lowest energy peak in the $||\mathbf{a}$ polarized spectrum can be attributed to positive correlations between fluctuations in $t_{e}$ and $t_{h}$.
Indeed, our mixed quantum-classical mapping approach predicts that the fluctuations in $t_{e}$ and $t_{h}$ are in fact positively correlated~(see SM).

We note that the simple dimer model described above cannot explain the nonuniform broadening observed in the TAT spectrum, or differences in broadening of different peaks in the $||\mathbf{a}$-polarized or ${\perp} \mathbf{a}$-polarized pentacene spectrum.
Additional nuances arise when local vibronic coupling is considered as the phonons allow electronic states of different symmetry to mix.\cite{spano_vibronic_2011,spano_reclassifying_2007}
Moreover, disorder in the system breaks the periodicity of the lattice, which also allows electronic states of different symmetry to mix.
A complete description of nonuniform broadening due to nonlocal electron phonon coupling requires an in-depth analysis of the Hamiltonian in Eq.~\ref{eq:ham}.
This should be the subject of future investigation, but is beyond the scope of the current work.

While selective line broadening represents an interesting effect of nonlocal electron-phonon coupling, the overall effect on the absorption spectrum is relatively minor~(Fig.~\ref{fig:PENTspec}).
As was the case for TAT~(Fig.~\ref{fig:TATspec}a), we find that for pentacene, the width of the coupling distributions are indeed quite large, and on the same order as their means~(see SM), in agreement with the results of Troisi and coworkers.\cite{troisi_dynamics_2006,arago_dynamics_2015}
Nevertheless, the spectral line shape is relatively unaffected by these wide distributions.

\section{Conclusion}
We describe an approach that efficiently incorporates nonlocal electron-phonon coupling of arbitrary form into model Hamiltonians for studying the photophysics and transport properties of organic crystals.
Our approach is inspired by mixed quantum-classical methods that are common in theoretical vibrational spectroscopy, and relies on an interpolation map for fast look-up of precomputed electronic couplings.
We apply this approach to study the absorption spectroscopy of two organic crystals with important applications in semiconductor devices.
We find that, even though the electronic couplings fluctuate over several hundred~\cm{}, the effect on the absorption spectrum is minimal and largely manifests as an increase in the absorption line width.
This explains how previous work, which often uses phenomenological line broadening parameters fit to experiment, has been able to accurately model the absorption spectra of organic crystals without accounting for nonlocal electron-phonon coupling.
The effects of nonlocal electron-phonon coupling cannot be entirely captured through an increased line broadening parameter, however, as we find that the line broadening is not uniform across the spectrum and that different peaks broaden to different extents.
Importantly, this explains, for the first time, the different line widths observed in the upper and lower Davydov components of the pentacene spectrum.
Using a model dimer Hamiltonian, we attribute the different line widths in pentacene to correlations between the fluctuations in CT couplings induced by nonlocal electron-phonon coupling. 

There are several possibilities for extending the approach presented here.
For example, maps of the time-independent quantities in Eq. \ref{eq:ham} could be generated to include fluctuations in those quantities in the model.
Other possibilities include generating the coupling maps on-the-fly over the course of the MD simulation until sufficient coverage of the sampled space has been obtained.
Machine learning approaches may also provide an avenue towards more accurate maps and towards relaxing the rigid-body approximation.\cite{kananenka_machine_2019,jackson_efficient_2019,jackson_electronic_2019, kraemer_charge_2020}
Finally, our mapping approach could be straightforwardly extended to include coupling to triplet excitons to investigate the effects of nonlocal electron-phonon coupling on singlet fission.

While spectroscopy is an important experimental probe in organic crystals, and can elucidate the nature of the electronic and vibronic interactions in a system, these systems hold most promise in the semiconductor industry, where charge transport properties like carrier mobility are of utmost importance.
The time-dependent Hamiltonian that our method computes can be used with any of the standard approaches that currently exist to evaluate the charge transport properties of a material.\cite{bondarenko_comparison_2020}
We expect that this will permit a comprehensive understanding of the effects of nonlinearities in the nonlocal electron-phonon coupling on the electronic properties of organic crystals.

\clearpage
\section*{Supplementary Material}
The supplementary material is available upon request.
It contains details about the MD simulations, the mapping procedure, the charge-transfer integral maps, probability distributions for the pentacene fluctuations, average nearest-neighbor CT energies, statistics for the coupling fluctuations, the complete parameter set for the spectroscopic calculations, comparison of different ab-inito methods for calculating the CT couplings, information about the phenomenological line broadening parameter, details about the spectral analysis, and details of the dimer model.

\begin{acknowledgments}
This work was completed with resources provided by the Pritzker School of Molecular Engineering and Research Computing Center at the University of Chicago. 
\end{acknowledgments}

\section*{Author Information}
\subsection*{Corresponding Authors}
\noindent Steven E. Strong: StevenE.Strong@gmail.com\\
Nicholas J. Hestand: HestandN@evangel.edu\\

\bibliography{lib}

\begin{thebibliography}{84}
\bibitem [{a_h()}]{a_hestand_polarized_note}%
  \BibitemOpen
  \href@noop {} {}\bibinfo {note} {Although this is not documented in
  Ref.~\citenum{hestand_polarized_2015}, that work used a linewidth of
  $\Gamma=374~\mathrm{cm}^{-1}$ for what we call the $\mathbf{a}$-polarized
  spectrum\cite{hestand_polarized_note_axes} and $\Gamma=214~\mathrm{cm}^{-1}$
  for what we call the $\mathbf{b}$-polarized spectrum.}\BibitemShut {Stop}%
\end{thebibliography}

\end{document}